# How to build a campfire? Participatory modelling with justice

*Nynke van Uffelen, Sander ten Caat, Aarthi Sundaram, Annemiek de Looze, Marion Collewet, Eefje Cuppen, Igor Nikolic*

**Abstract**
Energy transition decision-making is pervaded by deep uncertainties. Computational models are helpful in addressing such uncertainties, as they can give insight into techno-economic complexity. However, models alone are insufficient, as part of the uncertainties involve justice dilemmas. Moral deliberation and reasoning are therefore crucial. Participatory modelling (PM) is a promising approach that combines computational models with deliberation and might therefore be suitable for exploring energy justice dilemmas. This paper proposes a novel PM approach, which we call *PM with justice*, that (a) explores *questions* of justice; (b) includes the perspectives of local stakeholders and relevant voices 'beyond' the local case; and (c) facilitates dialogue and critical reflection on stakeholders' epistemic and normative assumptions. Using the metaphor of a campfire, the process entails five phases, namely (1) identifying justice issues, (2) participatory modelling, (3) involving publics, (4) building a model, and (5) collective critical reflection. The design was developed and applied in the context of the heat transition in a neighbourhood in the city of Utrecht in the Netherlands. We found that the PM approach, including the campfire metaphor, was useful in integrating justice and techno-economic considerations in collective reflection and deliberation on choice options for the heat transition in this neighbourhood. As such, PM with justice is a promising approach to address deep normative uncertainties in energy transitions.

**Keywords**
Energy justice, participatory modelling, energy transition, Participatory Value Evaluation, energy models

## 1. Introduction

Energy transitions entail thorough changes in complex sociotechnical systems, and thus decision-making is both technically and ethically challenging. Choice options generally have technological, economic, social, and ethical consequences and are pervaded with deep uncertainties regarding the ecological, economic and societal effects of options. Deep uncertainties are often addressed with the help of computational models (Süsser et al., 2021). Through calculations, models can explore the effects of choices, interventions and scenarios on, for example, the economy, $CO_2$-emissions, and other relevant metrics.

Although models are particularly apt to explore deep uncertainties, it is less clear if and how they can explore ethical issues in energy transitions. Energy transitions are pervaded with questions of justice, as choice options may benefit some groups whilst negatively impacting others. What the most just choice option is, is not always straightforward – in other words, there are normative uncertainties about energy justice (Van Uffelen et al., 2024). *Normative uncertainties* can be understood as uncertainty about what is ‘just’, as there are multiple - or no - options that are ethically justifiable (Taebi et al., 2020). For instance, urban heat transitions depend on technological innovations and constraints, and the (uncertain) impacts of different policies and technologies may give rise to energy justice dilemmas such as: how to weigh local inequalities and dissatisfaction against the benefits of a quick transition globally? Addressing deep uncertainties requires input from computational models, due to the immense techno-economic complexity – however, such models in themselves cannot address normative uncertainties (for an overview, see Sundaram et al., 2024, 2026).

*Participatory modelling* (PM) seems a promising approach to support energy transition decision-making pervaded with deep normative uncertainties. PM processes generally share two core elements, namely: (1) participation of stakeholders in the modelling process; and (2) knowledge sharing, collaboration, deliberation, and learning amongst involved stakeholders. The inclusion of a diversity of perspectives can aid in making the model more representative of reality and in making the model meaningful and more trustworthy to more stakeholders (Campo et al., 2010). When policymakers are involved in making a model, it will most likely be better suited to the policy need and it can help policymakers to understand what a model can and cannot do (Süsser et al., 2021). Moreover, by incorporating stakeholder knowledge and perspectives, models may be more relevant to the specific policy context (McGookin et al., 2024; Voinov et al., 2018). The exchange of knowledge and perspectives within PM processes may facilitate collaborative learning (Voinov & Bousquet, 2010), which may lead to real-world changes. As such, PM - if done right - has the potential to “address complex landscape and environmental problems like climate change, environmental *injustice*, and sustainable resource management” (Zellner, 2024, italics added).

Although existing PM approaches are increasingly developed and applied to incorporate (local) stakeholder knowledge in energy transitions (Heaslip & Fahy, 2018; Krzywoszynska et al., 2016; McKenna et al., 2018), they do not explicitly deal with normative uncertainties, specifically, questions of energy justice. It is wrong to assume that merely the inclusion of stakeholders in modelling processes makes it ‘procedurally just’ and that decisions based on such model inputs are then ‘legitimate’. After all, that would imply that what is accepted or preferred by the included stakeholders equals what is just. However, 'ethical acceptability’ cannot be equated with ‘social acceptance’ (Taebi, 2017), nor can 'justice’ be derived from the sum of individual preferences (Pesch & van Uffelen, 2024). This is because people's preferences may be based on false information, or motivated by interests, and more importantly, because certain voices may remain invisible and unheard during the modelling process (van Uffelen & ten Caat, 2025). We therefore see a need for a PM approach that can explore deep uncertainties including normative uncertainties, and specifically questions of justice, in energy transitions. So, our research question becomes: What does a PM approach look like that can explore deep uncertainties, including justice questions?

In this paper, we propose a PM approach that consists of five phases: (1) Identifying and prioritizing justice questions; (2) A PM workshop with key stakeholders; (3) Exploring citizens’ values; (4) Building a computational energy model; and (5) A reflexive workshop with key stakeholders. We developed and executed this PM process in the context of the energy transition within a neighbourhood in the city of Utrecht in the Netherlands, in collaboration with the municipality and a local neighbourhood cooperative.

The paper proceeds as follows. In section two, we elaborate on the concepts of energy justice and participatory modelling, and we argue for the need of a PM approach that allows for exploring justice questions. In the third section, we describe the PM approach. Section four describes how we executed the PM approach in the heat transition in Utrecht in the Netherlands. In the fifth section, we discuss our findings and reflect on the process and the approach. Section six concludes.

## 2. Energy justice and participatory modelling

### *2.1 Energy justice*

Three themes or tenets have come to be at the core of energy justice: distributive, procedural and recognition justice (McCauley et al., 2013; Santos Ayllón, 2025). These tenets pertain to the distribution of burdens and benefits, the decision-making procedures, and the recognition of actors through law, status order, and love (Van Uffelen, 2022). Whereas the three tenets have come to dominate in the energy justice literature, their use and their definitions vary (van Uffelen et al., 2024; Wood, 2023). Scholars employ the three tenets for instance to understand how diverse actors perceive justice. The tenets are also used to study whether energy policies and projects could be deemed just, or to formulate policy recommendations for more just energy systems. Such evaluative and normative use of the tenets builds on substantiated ideas of justice. Within energy justice research, such views of justice are often based in liberalism (Santos Ayllón, 2025). Although there is increasing attention to alternative perspectives (e.g. Dunlap & Tornel, 2023; Rainard et al., 2025; Sankaran & McIntyre-Mills, 2022; Sovacool et al., 2023; Tafon et al., 2023), these approaches engage only to a limited extent with the idea that there might be multiple morally defensible views of justice. As such, it is crucial to acknowledge that there may be multiple clashing conceptions of justice that are defensible in a certain context, a phenomenon that can be labelled as normative uncertainty (Taebi 2019; Van Uffelen et al. 2024). Justice, then, is a situated, contested and dynamic concept (de Looze et al. 2024), and which conception of justice is most suitable in a specific context should be deliberated and argued for.

### *2.2 Participatory modelling (PM)*

PM can be defined as "a collaborative approach to formalize shared representations of a problem using a wide range of qualitative and quantitative modelling techniques" (Zellner, 2024). Most PM approaches share two core elements, namely: (1) participation of stakeholders in the modelling process; and (2) knowledge sharing, collaboration, deliberation, and learning amongst different stakeholders. Current participatory modelling approaches are diverse and vary in the motivations for involving stakeholders, who is involved, and how.

There are instrumental, substantive and normative reasons for making modelling participatory (Fiorino, 1990). *Instrumental* reasons for participation reflect the idea that including stakeholders in knowledge creation through models would avoid resistance and distrust and consequently speed up energy transitions (Flacke & De Boer, 2017; McGookin et al., 2024). As for *substantive* reasons, some argue that stakeholder participation and knowledge in modelling processes generates better answers and increases the quality of model outputs (Nespeca et al., 2024; Prell et al., 2009; Smetschka & Gaube, 2020). Knowledge creation through traditional energy system models often privileges technical and expert knowledge, while leaving limited room for input from local stakeholders (Prell et al., 2009), affecting the quality of the models, as modelling "requires domain knowledge that spans several fields of expertise" (Nespeca et al., 2024). The inclusion of diverse perspectives can make models more representative of reality (Campo et al., 2010) and more policy relevant (McGookin et al., 2024; Voinov et al., 2018). Another substantive reason for PM is that it facilitates learning and collaboration. The exchange of knowledge and perspectives in PM generates shared understandings, which has the potential to lead to real-world changes (Duespohl et al., 2012; Krzywoszynska et al., 2016; Olabisi, 2010; Pahl-Wostl et al., 2008; Rodela, 2011; Voinov & Bousquet, 2010). In this, models function as boundary objects, bringing together stakeholders from different social worlds (Cuppen et al., 2021; Flacke & De Boer, 2017). Lastly, few authors give *normative* reasons for PM, namely that it is right or just to include stakeholders in the modelling process (Heaslip & Fahy, 2018; Krzywoszynska et al., 2016; Lonergan et al., 2023; Sundaram et al., 2026), thus presenting PM as intrinsically and ethically important.

PM processes also differ in who is engaged, how, when, and to what extent. Some PM processes include a diverse group of experts (Bernardo & D'Alessandro, 2019; Nespeca et al., 2024; Schinko et al., 2019), while others broaden the modelling actors towards, for instance, citizens (Heaslip & Fahy, 2018; Krzywoszynska et al., 2016; McKenna et al., 2018; Olabisi, 2010; Zellner, 2024). The methods to engage stakeholders also vary, ranging from consultation to active participation, with workshops, focus groups, interviews, and surveys among the most used (McGookin et al., 2024). Stakeholders can be engaged in different phases of the modelling process, varying from early-stage input gathering and problem framing to later-stage evaluation of model results or policy options (Sundaram et al., 2026; Meyer et al., 2025). There are also different levels of stakeholder engagement and participation: Van Bruggen et al. (2019) propose a typology, distinguishing between instrumental, representative and transformative PM approaches.

### *2.3 Participatory modelling with justice*

Some scholars aim to bridge (energy) systems modelling and justice (Hoffmann et al., 2026; Lonergan et al., 2023; Vågerö et al., 2024; Vågerö & Zeyringer, 2023) by, for example, quantifying principles of justice into the model logic, or by accounting for social factors (Krumm et al., 2022; Trutnevyte et al., 2019). Narratives for advocating PM approaches specifically, however, only rarely refer to (energy) justice, and mostly rely on the instrumental and substantive advantages of engaging stakeholders. We can identify two main exceptions to this, which we label (1) *justice as participatory modelling*, and (2) *participatory modelling for justic*e. We argue that neither approach is satisfactory and sufficiently acknowledges normative uncertainty, and we propose an alternative: *participatory modelling with justice.*

First, *justice as participatory modelling*. Based on normative reasons for PM, some authors call upon concerns of (procedural) justice to advocate PM as an approach in the context of sociotechnical energy transitions. For example, Lonergan (2023) states that public participation processes "support procedural justice in shaping the energy transition as well as increase the transparency and accountability of modelling exercises to the people who will bear the consequences of policy decisions made on the basis of the modelling results". Also, Sundaram et al. (2026) argue for PM approaches "to incorporate stakeholder input and operationalize justice". This view on PM can be labelled as *justice as participatory modelling*, indicating that concerns for (procedural) justice create an imperative to involve stakeholders in modelling, and doing so would make the modelling process more just. Although this argument is valid, it is not sufficient, because it remains unclear how PM can be leveraged to address deep normative uncertainties, including questions of justice.

Second, *participatory modelling for justice.* Based on substantive reasons for PM as explained in 2.2, it is suggested that PM generates better models, outcomes and results. However, PM does not automatically generate more *just* outcomes, because stakeholders have specific normative and epistemic assumptions, interests, and strategies, which shape their views on justice. As such, the sum of people's preferences cannot be equated with justice. Moreover, as PM is usually applied in specific local contexts, certain voices – especially voices from distant geographical regions, other species, and future generations, but also of marginalized people in the local present – may remain invisible and unheard during the modelling process (van Uffelen & ten Caat, 2025). The assumption that involving local stakeholders increases the legitimacy of the outcomes holds the danger of *participation washing*, in which outcomes are legitimized by referring only to stakeholder participation, while further ethical grounds are lacking (Sloane et al., 2020). Moreover, this assumption ignores that justice is pervaded with normative uncertainties. It takes at face value the dominant conception of justice that emerged from the participatory process, while alternative yet marginalized conceptions of justice may be more reasonable within the local context.

Acknowledging normative uncertainties and the risk of participation washing implies that the value of PM should not be sought in modelling that provides an *answer* to the question of what is a just choice option, but rather in its ability to facilitate the joint exploration of the question what justice means in the particular context, and the different perspectives and justice dilemmas that follow from that. We argue that, while PM can play a crucial role in navigating deep uncertainties in energy transitions, there is a lack of attention to how a PM approach can address normative uncertainties. As such, we propose a redesign. We call our approach *participatory modelling with justice*, treating 'justice' not as something to be achieved through participation only, but as a normatively uncertain concept that should be deliberated and discussed collaboratively in context, and for which a definite or clearcut answer might not exist. *Participatory modelling with justice* implies that questions of justice are central throughout the PM process, and its goal is to explore deep uncertainties, explicitly including justice questions, with stakeholders. In this, the model serves to explore justice issues, thus linking descriptive and normative realms, by asking questions such as: 'If option *x* were to have outcomes *y*, what would that imply for justice issue *z*?'

## 3. Designing a participatory modelling process with justice

### *3.1 Design rationale*

The *PM with justice* approach has three starting points. First, the goal is to use the model to collaboratively explore questions of justice, rather than to 'achieve' more just models that produce the answer to what is a just choice option. We consider justice a contested concept, as different defensible conceptions of justice may clash in specific situations. In other words, the PM treats justice as a question that should be tackled together: what could be the most just option in this context, and why? This question, in essence, is an applied normative and ethical-philosophical question. Acknowledging normative uncertainty and justice dilemmas means that a final and definite answer to this question may not be possible. This does not mean however that this question should be avoided altogether – on the contrary, it should be collectively explored. Such explorations "won't always yield decisive results. But that's the nature of our situation. We can't always be sure of things, in ethics or elsewhere. That shouldn't prevent us from trying to get it right and backing up our moral views with the best possible reasons" (Schafer-Landau, 2024, p. 13). As such, justice questions ought to be deliberated among stakeholders. In this, model outcomes do not directly point out which option is (un)just but rather which option should be considered food for thought and input for deliberation. This also implies that the model should be designed to explore the justice questions at hand – the form should follow the function, and not the other way around. It has to be noted, however, that justice questions cannot solely determine the type of model either, as model choice is also shaped by practical considerations such as available expertise and data, institutional support, and the legitimacy that can come from using a well-established model rather than developing a new one from scratch.

Second, the PM process should include the perspectives of key stakeholders, but also of relevant voices 'beyond' the local[1] case. In PM processes, the model scope, question, and purpose should be co-determined by key stakeholders, such as decision-makers and citizens. Consequently, the modelling process should be made intelligible to non-modellers. Still, justice issues and perspectives might not be sufficiently captured by merely involving local stakeholders. In the context of energy transitions, local, regional and (supra)national decisions may have impacts beyond the geographical boundaries and current generations, which actors embedded in local contexts may underestimate or overlook. This has implications for the identification of justice issues and justification of views, because 'what is just' for a region, species, or generation, may be in tension with 'what is just' for other regions, generations, or species. As such, the PM process should account for affected actors that are physically absent.

Third, the process must proactively elicit the epistemic and normative assumptions of all participants and facilitate critical reflection on premises and arguments. This is because questions of justice cannot simply be explored through a clash of perceptions and opinions. It is insufficient to simply collect the preferences and knowledge of citizens and take them at face value as true and valid premises in answering justice questions; instead, critical reflection ought to be directed at the reasonings behind them (e.g., Szetey et al., 2025). What is 'just' cannot be equated with what is socially accepted (Taebi, 2017). Similarly, it is crucial to critically interpret model outcomes and acknowledge the epistemic and normative assumptions leading to those outcomes.

### *3.2 A PM process that explores justice issues: building a campfire*

Based on these starting points, we use the metaphor of a campfire to explain the PM process. Just like a campfire, PM is a social process and a point around which people create and share stories. A campfire needs fuel, the wood, which is typically collected by everyone involved in building the fire. A good spot needs to be found, and the woodpile needs to be created and lit. People gather, share stories, watch at and interpret the flames. More wood can be added. And when that process is over, the fire is left to burn out. This metaphor seemed useful in explaining the PM process in an accessible way to stakeholders. In the PM process, stakeholders collectively build a campfire, in which the woodblocks represent epistemic and normative input – in other words, values,

[1] In our case study, 'local' means on the level of a neighbourhood. However, PM processes may also be leveraged on municipal, regional, or even (supra)national scales – as such, the term 'local' is relative.

perspectives, and knowledge – from the model, stakeholders, and affected actors. All gathered woodblocks are lit and collectively interpreted, with the aim of formulating a response to the justice question(s) at hand. We present the approach through this metaphor and propose five steps of the PM approach (see

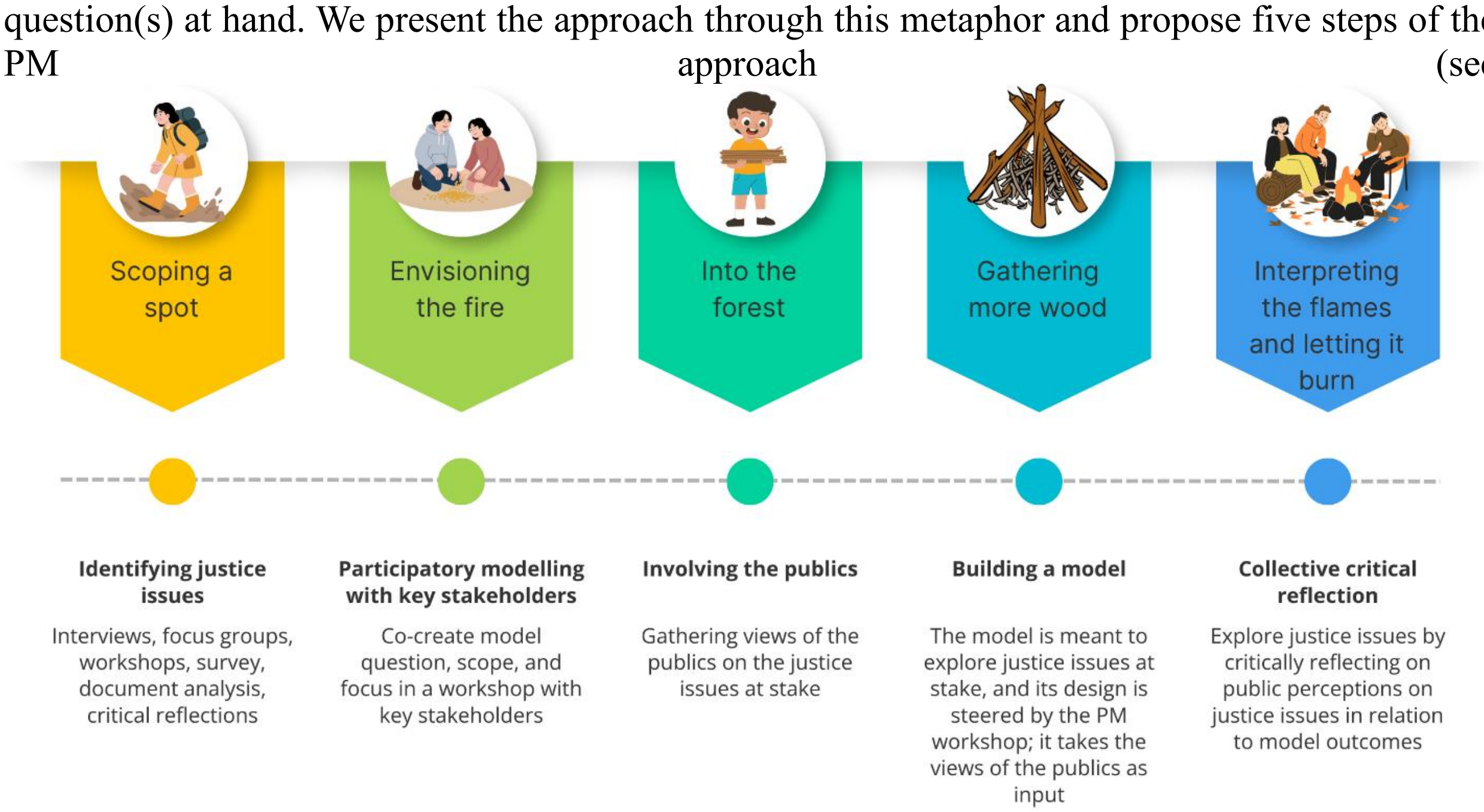

Figure 1), which we explain in the remainder of this section.

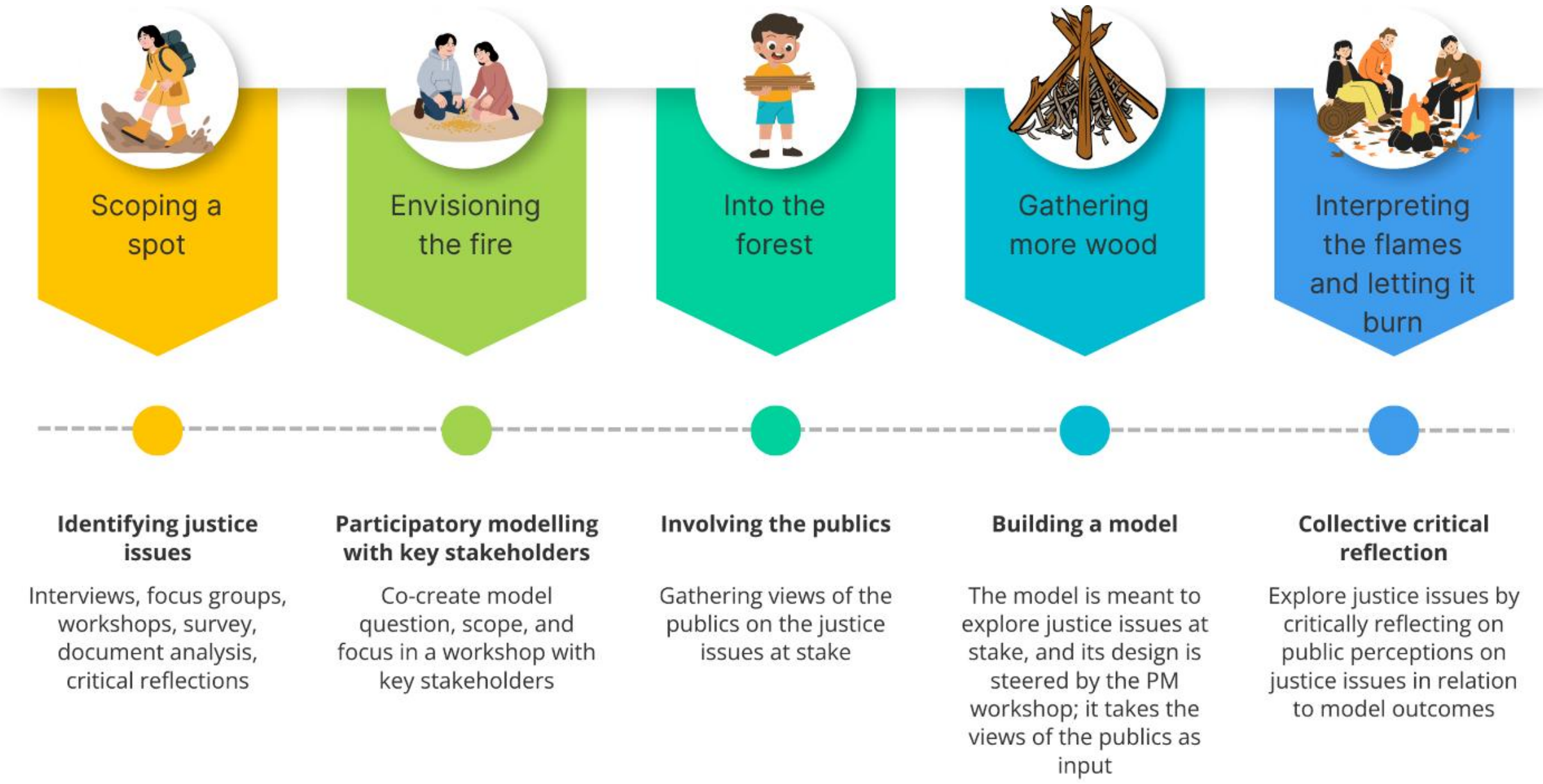

*Figure 1 - Five steps of the PM with justice approach, using the metaphor of building a campfire.*

*Phase 1. Scoping a spot: identifying justice issues*

Before starting to build a fire, it is important to scope the spot, or the context, in which to build the fire. As such, after specifying the case (which can be described as the pre-modelling phase, see

Sundaram et al. 2026), the first phase entails identifying the justice issues, trade-offs, questions, and dilemmas at stake. These can be wide-ranging and pertain to questions of procedural, distributive, restorative, or recognition justice. To do so, one may organize interaction with stakeholders through interviews, focus groups, workshops, surveys, or document analyses. In addition, the critical reflections of researchers are also valuable, as local stakeholders may emphasize justice issues 'here, now, and for us', with the danger of overlooking issues beyond the local. Specifically, actions and policies may generate local benefits yet may disadvantage actors elsewhere and in the future. As such, it is an important responsibility of the researchers to bring a wide range of justice issues on the radar of the PM process.

*Phase 2. Envisioning the campfire: participatory modelling with key stakeholders*

In the second phase, stakeholders collectively decide what the fire – in other words, the model – should look like, what it should (not) encompass, what kind of wood should be collected, and how it should be built. In this phase, key stakeholders co-create the focus of the PM process. With key stakeholders, we mean the actors that have decision-making agency, relevant actors that help shape the decision space, and (representatives of) affected stakeholders. Who the relevant actors are, can partly be derived from the first phase. These stakeholders can thoroughly steer the PM process, first in relation to the justice issues identified: which issue(s) do they recognize and find most important to focus on? Next, they co-shape the modelling question and the scope of the model, in short, they steer what the model should focus on. This process requires is most easily conducted with a select group of stakeholders through collective brainstorming in a workshop setting.

*Phase 3. Gathering wood: involving the perspectives of the publics*

Justice issues in local energy transitions are relevant to and affect a larger group of stakeholders, and as such, it is important to include societal actors in decision-making (Burke & Stephens, 2017). In PM processes, however, direct involvement of wider publics is often difficult due to high levels of abstraction, technological complexity, and knowledge prerequisites. Moreover, citizen representatives – for example through energy cooperatives – are rarely representative (e.g., Hanke & Lowitzsch, 2020; Boostani et al. 2024, 2026). As such, it is important to leverage methods to collect the views, perspectives and values of the wider local publics on the justice dilemmas at hand, beyond the small group of key stakeholders involved in the PM process. Possible methods are (online) surveys and discrete choice experiments. The outcomes of such processes can be seen as 'woodblocks' that are added to the fire, so that they can be taken into account in exploring the justice question(s) at hand.

*Phase 4. Building the campfire: building a model*

Next, the campfire should be built using the gathered wood, and according to the decisions made in the second phase. As such, the fourth phase focuses on building a model that can help to explore the justice issues, dilemmas, and trade-offs identified in the first phase. Depending on the justice issue(s) at hand, models can be leveraged to computationally explore possible outcomes, scenarios, dynamics, and trade-offs under different assumptions or uncertainties. The model conceptualisation is based on phase 2, in which stakeholders co-create the modelling question, scope, and requirements, and on input from the involvement of wider publics from phase 3. In our case, a pre-existing model formed the starting point for practical reasons discussed previously. However, the model was adapted, expanded, and used in accordance with the outcomes of previous phases. In a similar vein, choices about the spatial and temporal scope of the model(s), the relevant output parameters, and the types of dynamics or impacts to be represented are guided by stakeholder needs and public input in phase 2. Although the actual development of the model, including its implementation and coding, may be carried out primarily by the modeller, the aim is to ensure that key modelling decisions remain traceable to stakeholder input and involvement. Additional assumptions or technical choices should be explicitly documented and made available for discussion with stakeholders.

*Phase 5. Interpreting the flames and letting it burn: collective critical reflection*

Finally, the campfire is lit and people jointly watch and interpret the flames. Sticking to the metaphor, in this phase participants jointly reflect on the campfire, to see what insights the combination of campfire spot, structure and gathered wood provides, and what changes could further improve the flames. More precisely, stakeholders explore justice issues through a critical collective reflection exercise on (a) the model outcomes, which implies highlighting the assumptions made in the modelling process to avoid misinterpretation and reflect on underlying values and ideas; (b) the publics' views, as they may bring new viewpoints and insights but may also contain inconsistencies, invalid reasonings, or false information; and (c) the stakeholders' epistemic and normative assumptions and reasonings, for the same reason. As such, all gathered woodblocks - that is, one's own assumptions, publics' views, and model outcomes – should be collectively and critically scrutinized, as if the woodblocks were premises in a moral argument, and key stakeholders work together in drafting a response to the justice question(s) at hand. Finally, the flames will – and should – burn out. The model process and outcomes affect people's views, actions, and priorities, and as such, the model outcomes cannot reflect an objective truth or even make predictions. Participatory models do not reflect realities but shape them, and as such, they always invalidate themselves.

## 4. The energy transition in Utrecht Noordoost

The PM with justice approach was applied in the context of the energy transition in neighbourhood Noordoost in Utrecht, the Netherlands, between February 2025 and April 2026. In particular, the focus was on the transition of heating of buildings in that neighbourhood. The Dutch government has set the target that by 2050, all buildings in the Netherlands must be heated without natural gas. This transition to sustainable alternatives for heating houses and hot tap water is referred to as the heat transition. This case was chosen because of an upstarting and newly funded collaboration between the municipality and a local neighbourhood cooperative, supported by several other organizations, to accelerate the heat transition, without explicitly focussing on justice. This provided us with an opportunity to engage them in a PM process and bring in new dimensions to their project. Moreover, the relatively privileged character of this neighbourhood, based on demographic data[2], provided an interesting opportunity to explore justice tensions between different geographical scales, as decisions that may be just for the neighbourhood may lead to injustices in other parts of the city, country, or the globe. We conducted this research in the context of a research project on energy justice and modelling, which was led by an academic team and involved public and private partners in its consortium. These partners were experts in the domains of modelling, participation and energy systems. Some of them were involved in the PM process.

[2] https://opendata.cbs.nl/#/CBS/nl/dataset/85984NED/table?dl=D7F5A

### *4.1 Scoping a spot: justice issues in Utrecht Noordoost*

To identify justice issues, we conducted 12 semi-structured interviews with stakeholders directly involved in the neighbourhood's heat transition (3), local heat transition professionals (6) and with experts in the Dutch energy transition (3). The interviews focused on interviewees' views on the most important elements of local heat transitions, the most significant challenges, possibilities and issues in relation to technological solutions, and on the possible next steps, disagreements and choices to be made. Moreover, we conducted a document analysis, in which we studied policy documents on the neighbourhood's heat transition produced by the municipality and the neighbourhood cooperative. Lastly, in a brainstorming session, the research team, in collaboration with energy transition professionals, imagined possible additional justice issues. The identified issues reflected justice dilemmas (such as individual choice freedom of local inhabitants versus steering towards a quick transition, which would benefit a broad range of people and non-humans globally), potential unjust (distributive) consequences of choices (such as the order of neighbourhoods that will be decarbonized), and questions of inclusivity, participation, compensation, and ownership (see attachment 1).

### *4.2 Envisioning the fire: the PM workshop*

Next, we organized a PM workshop with key stakeholders to determine focus, scope, and goals of our collective endeavour. Following the justice issues identified in the first phase, key stakeholders were the municipality of Utrecht, the neighbourhood cooperative, and citizens. From the research consortium, three researchers were present, as well as four consortium partners, namely three participation experts who helped structure and moderate the workshop, and one modelling expert. Four actors from the municipality and three people from the cooperative were present, as well as three people from the grid operator, because they help shape the decision space; one energy modelling expert from a company that was a consortium partner in the research project; one[3] employee from an organizations involved in the pre-existing collaboration between the municipality and neighbourhood cooperative in their capacity as coordinators and mediators. The workshop was in-person, held in the neighbourhood, and took three hours.

In this setting, we first proposed and discussed a policy question, to identify the question(s) the decision-making actors wanted to start with, and their main interest and curiosity. The question that was proposed, was: "What heat technologies and sources can we best deploy in the neighbourhood, and which order of neighbourhoods should we maintain when phasing out natural gas?" To this question, the participants responded that the question was too technical, and that they would like to focus more explicitly on justice. As such, the question was transformed into: "What is the most just way to deploy heat technologies in the heat transition in Utrecht Noordoost (and where to start)?"

[3] Another person from another involved organisation was invited, but was ill on the day.

Next, we presented the justice issues and let everyone read, interpret, and rank them by putting stickers on the issues that they found most pressing or interesting to explore within this project (attachment 1). In a group discussion, we discussed the prioritisation, focussing on the interpretation of the issues, and why certain issues had (almost) no stickers, as we wanted to check whether these issues were interpreted by the participants as we intended, or if the participants did not recognize these issues.

After this, we co-created a modelling question, using the XLRM-framework (Lempert et al., 2003; Popper, 2019). We used the XLRM-framework because of its simplicity, as it focuses on the core elements that shape the model. The participants brainstormed about the external factors (X), policy levers (L) and the metrics (M) they want the model to explore the relations (R) between. Due to time constraints, the prioritization of the outcomes of this brainstorm was organized through a follow-up survey in Qualtrics (see attachment 2). In the last part of the workshop, participants discussed the modelling scope, i.e., the boundaries in terms of space, time, and social factors. In balancing the preferences of participants and modelling feasibility, it was decided to focus on the levers that the municipality and neighbourhoods have (and the decisions inhabitants can make), and to regard decisions made on European and national levels as external factors; to model the impact on the neighbourhood only; to restrict the timeframe to 2030, and where possible or applicable, 2050; and to distinguish between different social groups when modelling impacts where possible.

### *4.3 Gathering wood: the participatory value evaluation*

We conducted a Participatory Value Evaluation (PVE) to engage the publics in our PM process. PVE is an online survey-style participation tool meant to elicit people's values in relation to policy options. Due to its online nature, it promises a low barrier to participation, resulting in a more diverse and younger group of respondents compared to offline approaches (Mouter et al., 2021). PVE aims "to put citizens into the shoes of a policy maker" (Mouter et al., 2021, p.4) by confronting respondents with a set of realistic policy options, policy targets (e.g., $CO_2$-reduction), and constraints (e.g., available funds) (Mouter & Beumer, 2024). Participants can also motivate their choices, thus providing quantitative and qualitative insights into citizens' values.

The PVE was grounded in the identified justice issues (see attachment 3.6) and was co-designed with the municipality and neighbourhood cooperative. The researchers had 10 meetings on the questions, tone and style, timing, dissemination and analysis of the PVE, to ensure that the PVE fitted the needs of both the key stakeholders and the researchers.

The PVE consisted of general survey questions and two choice-tasks (attachment 3). To start, respondents answered survey questions on their demographics and general opinions on climate change and the heat transition. In the first choice-task, respondents were asked to divide 45 points over 9 statements, indicating their relative importance. The statements reflected a set of values respondents could prioritize in the heat transition that are all related to the justice issues identified in the first phase, namely participation, cost saving, individual choice freedom, expertise, a collective approach, being informed, mitigating climate change, and equality. This gave insight into participants' values under no constraints. In the second choice-task, respondents were asked to make choices along three dimensions by positioning sliders on three scales: (1)

Stimulate collective heating solutions or leave people to individually transition away from natural gas, (2) Divide costs of the heat transition over neighbourhoods, or let the inhabitants of each neighbourhood pay for the costs the transition in their area incurred, and (3) Minimise or maximise the budget allocated to the spatial embedding of heat technologies in the local neighbourhood. At the same time, they could see on their screen how their choices affected four outcomes: (1) End-user costs, (2) Inequality among neighbourhoods, (3) A pleasant environment in their own neighbourhood, and (4) Whether their neighbourhood had successfully transitioned away from natural gas by 2050. The choices and their effects relate to several justice issues and elements from the XLRM-brainstorm exercise. By confronting respondents with the consequences of their choices, we sought to study how participants' values were affected when faced with constraints.

The PVE was open to all citizens of Utrecht Noordoost, and was distributed through municipality channels, at four information evenings on the heat transition, door-to-door-flyering to increase respondents in less represented areas, and through advertising on social media and the newsletter by the neighbourhood cooperative. In total, 635 participants filled out the entire PVE. The respondents were not representative of the inhabitants of the neighbourhood: men, people over the age of 45, highly educated people, homeowners and retired people were overrepresented, despite our attempts to reach less represented groups.

We used linear regressions to study how participants' choices in the first and second choice-tasks correlate with their demographic characteristics and opinions about climate change and the heat transition. We also attempted to identify clusters of respondents based on their choices using a latent-class cluster analysis, but this did not yield clear-cut clusters. Moreover, two researchers qualitatively analysed the explanation from participants inductively.

As for the quantitative results, an overwhelming majority (86.3%) felt somewhat or very concerned about climate change and 60.4% found the heat transition somewhat or very important. Both elements correlated with a preference for collective solutions. Moreover, valuing individual choice freedom correlated with a preference for individual solutions. Notably, participants who awarded most points to low end-user costs in the first choice-task still generally chose for individual solutions in the second choice-task, even though that generated higher costs. A reason for this may be found in the qualitative explanations: respondents seemed to heavily rely on personal ideals, such as 'individual choice-freedom' to choose either for a collective or individual approach to the heat transition or 'togetherness' (the idea that we ought to do the heat transition together, not individually), and these ideals seemed to trump cost concerns when filling in the survey.

### *4.4 Building the campfire: modelling extremes*

Given feasibility constraints and stakeholder preferences, the model focused on one justice issue, namely 'individual choice freedom versus steering towards a quick transition'.[4] To explore this dilemma, we designed three scenarios, namely (1) individual freedom: everyone chooses themselves when to transition and by using what heating solution, (2) steering towards collective heat grids, and (3) steering towards all-electric heat solutions. These scenarios can be considered as extremes, allowing for a clear comparison and evaluation of the desirability of options, given that the justice issue at hand is a trade-off between two opposing values. Per scenario, we estimated

[4] Although the model could do more, we focused on this justice issue, because the stakeholders in the PM process did not have the time to explore multiple justice issues.

(a) the total investment costs, that is, the sum of public and private costs, including expansion of the electricity grid, heat pumps, and the necessary infrastructure for heat grids, and (b) its impact on the above-ground public space ($m^2$). Although the numbers the model produces can only be interpreted as rough estimations, the model is sufficiently fine-grained to compare the three extreme scenarios, allowing for conclusions such as 'scenario X will take up more public space than scenario Y'.

For the model, we used Hestia as the starting point, a decision that was based on stakeholder input during the second phase. Hestia is a simulation model developed by the Dutch Environmental Agency (PBL) and an applied research institute (TNO) to explore how the energetic qualities of the Dutch residential building stock change in response to different policy scenarios (van der Molen et al., 2023). Exploring these scenarios through the model required the ability to explore impacts of resident-level choices of different heating technologies, which could not be done with Hestia as it was, as neighbourhoods are the smallest spatial unit of analysis. We thus developed an additional microsimulation model that interfaced with Hestia, such that for each scenario, the model assigned the corresponding heat technology to individual dwellings. For instance, in the scenario steering towards a rapid transition, heat grid connections were assigned to relevant dwellings, overriding their existing heating installations. In the all-electric and individual freedom scenario, dwellings that opted for individual solutions were assigned either air-source or ground-source heat pumps. Based on these assignments, the model recalculated energy demand per energy source and demand category (e.g., space heating, hot water). For electricity, the resulting demand was used to estimate changes in peak demand and the electricity-grid reinforcements or expansions required to accommodate it. For heat grids, the model estimated the required expansion of the heat network along the existing street network. More details on how the calculations were made can be found in attachment 4.1.

In the individual freedom scenario, the model took people's preferred heat solutions as indicated in the PVE – that is, heat pumps (both air-source and ground-source) or collective heat grids – at face value and extrapolated them within each of the 11 districts within the neighbourhood, leading to a mix of both heat pumps and heat grid connections. In this, we assumed that every individual must go off gas eventually, and as such, people who refused the heat transition altogether were assumed to have an equal chance to prefer heat pumps and heat grids. The model calculated the estimated costs and the impact on public space in the neighbourhood if every individual were given the freedom to follow their individual preferences. In the scenario of steering towards collective heat grids, we calculated the consequences of connecting 90%[5] of all households within each area within the neighbourhood to a heat grid. Lastly, in the scenario of steering towards all-electric solutions, we estimated the spatial and financial effects if every household were to adopt heat pumps.

The model outcomes (attachment 4.2) show that following people's individual preferences would be the most expensive scenario in terms of total investment costs, while steering them towards collective heat grids would be the cheapest. This gives a trade-off between individual choice-freedom and costs, and the PVE results suggest that both are deeply valued by both stakeholders and citizens. Moreover, the model estimates that steering towards collective heat grids has significant impact in the above-ground public space, followed by the scenario of individual freedom. This constitutes a dilemma between heat grids, which is something most key stakeholders and PVE-participants seemed to prefer, and the availability of scarce and valued public space.

---

[5] We assumed a 90% rather than 100% connection rate, as even under strong steering, complete uptake of a collective heat grid was considered unrealistic.

*4.5 Interpreting the flames and letting it burn: the campfire workshop*

In a final 'campfire' workshop, seven stakeholders reassembled to critically reflect on the PVE and model outcomes (as well as six researchers, two consortium partners who were participation experts that helped shape the workshop, and two consortium partners that were modelling experts from two different companies). The intention was to reassemble the same group of participants as in the first workshop. However, due to practical reasons, there were some changes, for example, a stakeholder from the distribution system operator was missing. The focus was on the highest ranked justice issue, namely 'individual freedom versus steering towards a quicker transition'. We sticked a poster with this issue on the wall, with individual freedom on the left and steering on the right, with a moveable arrow in between, designed to adapt during the workshop through deliberation. The workshop was framed as a 'campfire session', and in the middle of the room was a 'campfire' (with LED light) consisting of unwritten (foam) woodblocks (for an impression see Figure 2). REF _Ref235542047 \h Participants were told that more woodblocks would be placed there during the workshop with gathered epistemic and normative assumptions and information, and that they were free to (re)move and edit them throughout the workshop.

After a short recap, the participants filled in a poll about their own epistemic and normative assumptions (see attachment 5). Remarkably, 7/12 participants estimated that heat pumps would be the cheapest solution for the neighbourhood, while two thought that heat grids would be cheaper. Still, most participants preferred collective solutions, mostly to avoid grid congestion, because the heat transition is a collective problem requiring a collective solution, and because it would accelerate decarbonisation. In relation to the dilemma at hand, participants generally leaned slightly towards "steering towards a quicker transition", with an average response of 6,42/10, the arrow was placed accordingly in the continuum on the wall in the workshop space. Some woodblocks describing normative and epistemic assumptions that emerged from the poll, and also earlier in the process, were placed on the fire (for an overview of all woodblocks placed during the workshop, see attachment 6).

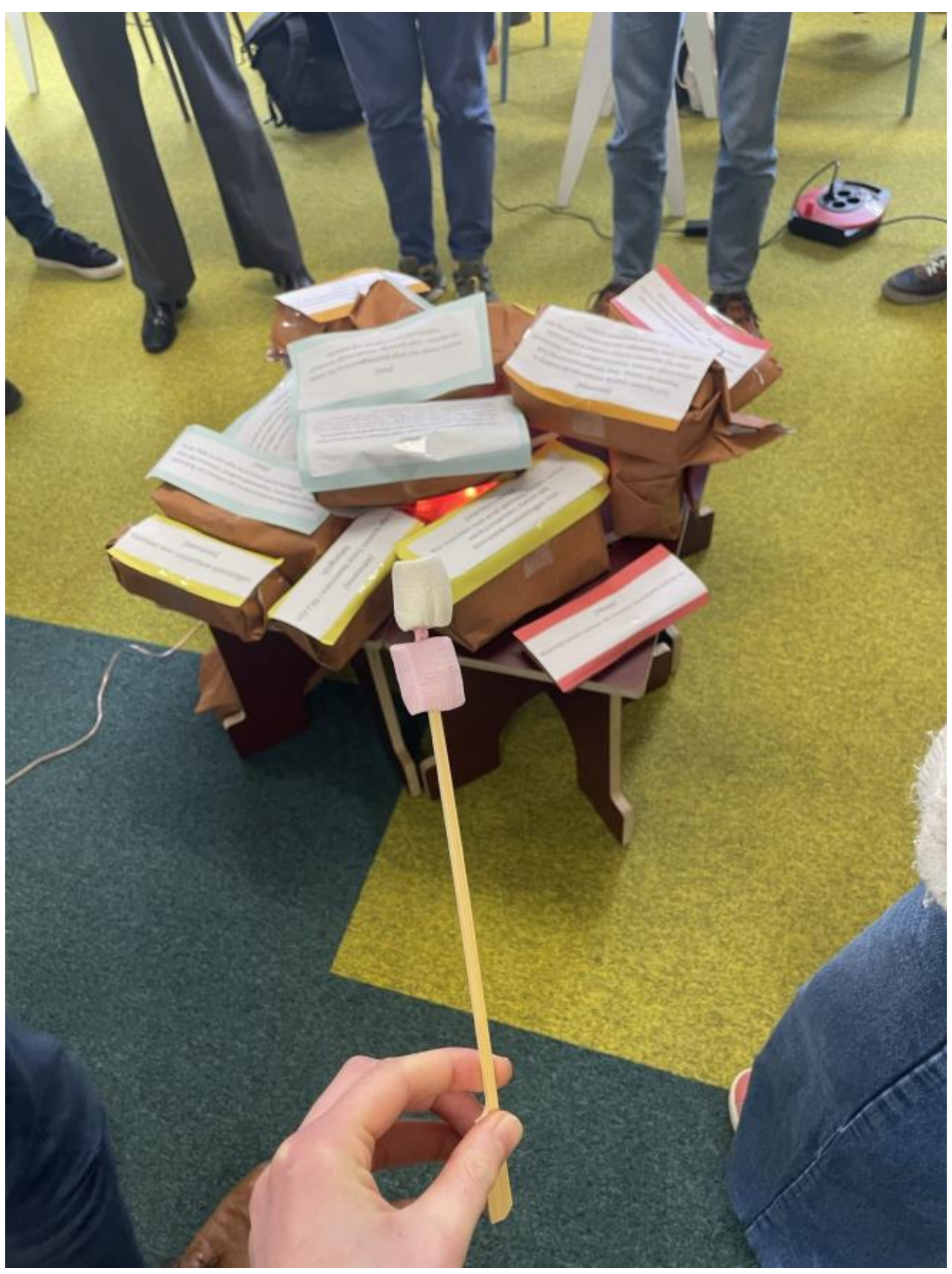

*Figure 2 - An impression of the 'campfire' with woodblocks, representing normative and epistemic assumptions, PVE and model results, during the campfire workshop in Utrecht Noordoost.*

Second, the researchers presented the first PVE results, pertaining to the demographics of respondents (and hence the lack of representativeness of the survey), the results from the survey questions and the first choice-task. Woodblocks that contained this information were added to the fire (for an overview of all woodblocks that were added throughout the workshop, see attachment 6). Participants engaged in a discussion about the results. During this discussion, the stakeholders started to lean more towards steering towards a quick transition and hence a collective solution. Some stakeholders pointed out that steering increases the least well-off citizens' freedom, because if a lot of people were to buy heat pumps, collective heating would become more expensive and thus no longer a real choice-option. Moreover, a quick transition would benefit the least well-off if combined with a variety of social goals, and as such, steering was considered just. Lastly, stakeholders noticed that it was mostly the climate sceptics in the PVE who very much valued their individual freedom.

Next, the model outcomes were presented and new woodblocks were added. In the discussion that followed, participants aimed to interpret and understand the model results. In relation to the estimated costs, participants questioned the accuracy, validity and scope of the model (as it excludes end-user costs). This led to a more fundamental discussion on the usefulness of modelling costs as input in decision-making, because cost estimations were considered "very unstable". Later in the workshop, the woodblock on estimated costs was removed from the campfire, as it was considered too weak to draw conclusions from.[6] Debate also arose in relation to the estimated impact on the public space, which allowed participants to formulate a new trade-off: "So the cheapest option requires the most public space, so that is a dilemma." Participants posed the question of how much public space is worth to them, and how much they are willing to sacrifice for it. During the discussions, participants moved the arrow further to the right, becoming increasingly in favour of steering for a quick transition, mainly because of the abundance of physical, financial and feasibility constraints to decision-making options, and as citizens "simply cannot oversee" these complexities and constraints.

Next, the results from the PVE second choice-task were presented and added to the campfire. For example, we showed that costs were not always the deciding factor for PVE participants when making choices, as for some people, individual choice-freedom carried even greater weight.

Finally, two 'zooming out' woodblocks were also inserted, pertaining to the global unequal impact of and responsibility in causing climate change, and to the local environmental impact of mining for critical materials for energy technologies, namely: "'Self-determination' privileges the interests of the citizens in the neighbourhood. However, climate change predominantly impacts people, non-human animals and nature outside the neighbourhood. In general, the Global North contributed more to climate change than the Global South", and "Heat pumps require many resources and materials that rely on harmful mining practices, more so than heat grids".

The discussion prompt was as follows: if we consider all woodblocks that are gathered so far, what does this imply for the central justice dilemma? When discussing the new woodblocks, participants moved the arrow even more to the 'steering for a quick transition' side. For example, one participant said that "when taking into account the bigger picture, I see myself moving more to the right", referring to the impact of mining resources. Another participant stressed the importance of biodiversity in the neighbourhood, to which another replied: "More agency from the side of the government, as they make a more integral reasoning than individuals. (…) Governments can consider these things, more than individuals will do, I think." Given the premise that individuals have a narrower view and limited knowledge than the municipality and the neighbourhood cooperation, the stakeholders found a certain amount of steering justified.

Overall, throughout the campfire workshop, the arrow in the continuum "individual freedom versus steering towards a quick transition" shifted significantly to the steering-side.

## 5. Discussion, evaluation and reflection

In 5.1, we discuss to what extent the first design rationale was met, namely approaching energy justice as a question to be explored. In 5.2, we relate the approach to the second and third starting points, namely the inclusion of perspectives of local key stakeholders, citizens, and actors 'beyond' the local case, and critical reflection on one's own epistemic and normative assumptions. In 5.3, we reflect on the role the model took within the process, and its relation to normative uncertainty.

### *5.1 PM with justice: starting from justice questions*

[6] Interestingly, in the debrief, almost all participants argued that cost estimations, although uncertain, were still highly relevant in considering which heat solution to strive for.

The PM approach intended to consider energy justice not as an outcome to be achieved through participatory modelling, but as a question to be explored collectively. This was achieved by explicitly centring justice questions throughout the process from the get-go, and through the metaphor of a campfire, in which the woodblocks resembled relevant input to explore the justice questions and dilemmas at hand. The participants reflected positively on the use of the metaphor of the campfire, stating that it helped them put together pieces of information on citizens' values, implications of actions as described by the model, and their own normative and epistemic assumptions, in relation to the justice question. As such, we conclude that it is certainly possible to have in-depth and fruitful conversations with stakeholders on complex questions of energy justice and computational models, pervaded with epistemic and normative uncertainties, and that the metaphor of a campfire was helpful to create a clear structure for doing so.

Interestingly, at the end of the campfire workshop, participants flagged that they did not fully understand the link between the process and 'justice'. This may be considered problematic, yet it could also be seen as an asset. Justice is both a folk term as well as a complex academic concept, and participants did not feel the discussions were too academic, abstract or far-fetched, but concrete and related to their core concerns. So, whereas we introduced the process with the notion of energy justice, and justice was also central in the modelling question, we deliberately chose to talk about concrete choices and dilemmas for decision making in the remainder of the process. Explicitly stressing the use of the concept 'justice' may have made it more clear for participants that these choices and dilemmas are, in fact, justice questions, but would have needlessly confused the process and could even have been counterproductive.

The PM process did not result in a clear outcome or policy approach – which was also not its intention. Some participants found this frustrating yet appreciated gaining deeper insights and the confidence to proceed. For example, the neighbourhood cooperative mentioned that the process helped to get things moving, forcing the municipality and the neighbourhood cooperative to continuously discuss and get on the same page, and to gather more information to start making decisions. The municipality added that the PM process helped make clear their responsibility to act. As the arrow significantly shifted towards steering towards a quick transition, the municipality felt strengthened in their mandate in the heat transition.

To further improve the quality of the deliberation, we recommend taking more time for the campfire workshop. Moreover, in the first two parts of the workshop, participants remained seated, even though they were invited to stand up and physically engage with the wood blocks in the middle of the room. In the third part, participants were encouraged again to move through the room and (re)move and (re)organize the woodblocks, which benefitted the discussion. Given the value of making use of the materiality of the campfire, we recommend creating a setting in which stakeholders physically engage with the content on the woodblocks.

### *5.2 Inclusion*

In our design rationale, we described the importance of including both the publics and voices and perspectives 'beyond the here and now' in the PM process. We involved the local publics through the PVE, and their perspectives led to critical reflection on the epistemic and normative assumptions held by the key stakeholders, thus influencing decision-making (Hendriks, 2010). However, it must be noted that the PVE was not representative. In contrast to Mouter et al. (2021) and Mouter & Beumer (2024), we did not find that the nature of the PVE tool allowed us to include a larger share of young people, lower-educated citizens, or a 'silent majority' beyond the usual suspects. Only 49,6% of all people who started the PVE finished the two choice-tasks, and some key stakeholders argued that the questions were difficult and lengthy. It is important to keep striving for representativeness, possibly through using mixed methods, yet simultaneously include reflexiveness in the PM process on hidden perspectives, as it may be an impossible ideal to achieve in practice.

To include perspectives of unrepresented actors beyond the local case, we followed Dryzek & Niemeyer (2008), Cuppen (2012) and Van Uffelen & Ten Caat (2025) in aiming for discursive inclusion and a reframing of the dominant, localized policy debate. This was the purpose of the two 'zoom-out woodblocks' in the campfire workshop. Although the blocks were limited in number and cannot substitute for actors expressing lived experiences, they did bring in global and non-anthropocentric perspectives. Both woodblocks were related back to the initial justice issue and served as further justification for steering towards a quick transition.

Another inclusion challenge came up in relation to the interests of stakeholders and the scope of the case. The neighbourhood cooperative had an interest in focusing on (potential) injustices for neighbourhood residents. In contrast, the municipality focussed on the interests of all citizens in the city, and they were aware that some decisions that may be good for Noordoost may negatively affect other neighbourhoods. Still, they were eager to collaborate with the neighbourhood cooperative and thus often prioritized their desires. In co-shaping the PVE, key stakeholders actively advocated against opening the survey to the city as a whole and insisted on keeping the focus of the survey locally. Consequently, certain research aims – namely, studying the effects of confronting local stakeholders with the perspectives of other neighbourhoods – could not be achieved. This raised several questions: which stakeholders' preferences should be decisive when determining the model scope? For example, what if local stakeholders were to dismiss or give low priority to justice issues that pertain to potential injustices beyond their concern? These concerns resonate with findings by Lim et al. (2023), who point out that PM opens the modelling process to political influences. Overruling the priorities and concerns of the stakeholders, however, may generate resistance and endanger the collaboration. As researchers, we found it crucial – and even a moral responsibility – to keep emphasizing unjust effects of decisions beyond the local, and to challenge the perspectives of local actors by bringing in justice issues and perspectives from absent actors. We addressed this challenge by focusing on the highest ranked justice issue(s), but where possible, the PVE and the model also incorporated elements that allowed stakeholders or other researchers to explore the 'marginalized' justice issues elsewhere and in the future.

### *5.3 The role of the computational model*

The goal of the PM with justice approach was to explore deep uncertainties, including questions of justice. This goal differs from other PM approaches as it explicitly focuses on normative aspects of decision-making jointly with technical, economic, and social aspects. As such, the role and functions of the computational model in this process are worth considering.

To our knowledge, the role of models to explore deep uncertainties is not yet described in the literature. Edmonds et al. (2019), for instance, described seven purposes of modelling, namely prediction, explanation, description, theoretical exposition, illustration, analogy, and social learning. Also, most PM literature focuses on social learning (ADD REFS). As such, we propose an additional purpose models can have, namely *normative exploration*, that is, the model facilitates the collective exploration of a moral, ethical, or justice question, dilemma, or trade-off.

The model fulfilled this purpose in various ways. For one, the model outcomes served as 'premises' in the conversation, giving rise to reflections such as: 'If option *x* were to have outcomes *y*, what would that imply for justice issue *z*?' The model helped make abstract justice dilemmas more tangible by showing what is at stake. As such, the model formed the bridge between the technical world – the world of materiality and physical constraints – and the ethical realm, which gives meaning and value. In a way, the model functioned as a boundary object (Cuppen et al., 2021; Jakku & Thorburn, 2010), as it facilitated stakeholders from different social worlds to jointly reflect and deliberate on an issue. Moreover, both the PM process and outcomes revealed core values of stakeholders. For example, during the campfire workshop, the estimated costs were heavily contested, which generated interesting reflections on the dominant role of cost estimations in heat transition planning, and the values and interests of key stakeholders, including possible moral blind spots. Lastly, because of the model's ability to ground normative issues in the material world, the model also served to reveal additional justice issues. Specifically, during the campfire workshop, the spatial impact of different scenarios made participants aware of another justice trade-off, namely of public space versus values such as costs and collective action.

## 6. Conclusions

In this paper, we proposed a novel PM approach, 'PM with justice', in which stakeholders collectively explore justice questions, issues, or dilemmas. Instead of striving for 'just model outcomes', this process acknowledges that justice is pervaded by normative uncertainties, and it centres the question: ' what could be the most just option in this context, and why?' Addressing this question requires a participatory process, input from computational models, input from stakeholders and publics, and critical reflection on one's own epistemic and normative assumptions.

We illustrated the design rationale and steps of the PM with justice approach, using the metaphor of building a campfire, and we showed how we applied the approach in the context of the heat transition in Utrecht Noordoost in the Netherlands. From this, we conclude that the campfire metaphor was effective in organizing reflection on a justice issue in the heat transition, bridging techno-economic and ethical realms. Moreover, we reflected on the problem of inclusion and argued that researchers have a moral responsibility to keep challenging the local focus of PM processes by 'zooming out' and organizing critical reflection. Based on all this, we conclude that PM can be used for the purpose of normative exploration.

We hope to inspire researchers, modellers and policymakers to acknowledge normative dimensions in (research, modelling and policy) questions on energy transitions, and to address them through the approach of 'PM with justice'. In this, we stress that engaging stakeholders in modelling processes does not guarantee more just outcomes, as justice is pervaded by normative uncertainties. Future research may be directed at further developing this approach. Moreover, it may be worthwhile to explore 'design with justice' as a general modelling and engineering approach. In contrast to 'design for justice' (e.g. Costanza-Chock, 2020), which aims to design technologies for just outcomes, 'design with justice' treats justice as normatively uncertain, and design serves to facilitate collective reflection on justice issues, questions, and dilemmas.

**Code availability statement**

The model code and data underlying the findings described in the manuscript are available under an appropriate Open Source license. The model code and dependent data are deposited publicly at [link].

**Acknowledgements**

Many thanks to the neighbourhood cooperation, the municipality of Utrecht, and the other key stakeholders in the case for their time, commitment, and collaboration. We also want to thank the consortium partners involved in the JustETrans project who helped prepare and support the workshops.

**Funding**

This paper was written as part of the JustETrans project (www.justetrans.nl), which is funded by the Dutch Research Council (NWO) under projectnumber KICH1.ED03.20.002.